\documentstyle[prd,aps,epsf,floats]{revtex} 
\draft

\begin{document}
\twocolumn[\hsize\textwidth\columnwidth\hsize\csname
@twocolumnfalse\endcsname
\title{
\hbox to\hsize{\normalsize March 1996\hfil FERMILAB-Pub-96/058-A}
\hbox to\hsize{\normalsize Submitted to Phys.~Rev.~D \hfil 
MPI-PTh/95-130}
\hbox to\hsize{\normalsize \hfil SFB-375/68}
\hbox to\hsize{\normalsize \hfil BA-96-09}
\vskip0.9cm
Supernova Neutrino Scattering Rates Reduced by Nucleon Spin 
Fluctuations: Perturbative Limit}
\author{Georg Raffelt}
\address{Max-Planck-Institut f\"ur Physik,
F\"ohringer Ring 6, 80805 M\"unchen, Germany}
\author{David Seckel}
\address{Bartol Research Institute,
University of Delaware, Newark, Delaware 19716}
\author{G\"unter Sigl}
\address{Department of Astronomy and Astrophysics,
The University of Chicago,\\ Chicago, Illinois 60637-1433\\
and NASA/Fermilab Astrophysics Center\\
Fermi National Accelerator Laboratory, Batavia, IL 60510-0500}
\date{\today}
\maketitle
\begin{abstract}
In a nuclear medium, spin-dependent forces cause the nucleon spins to
fluctuate with a rate $\Gamma_\sigma$. We have previously shown that
as a consequence the effective axial-current neutrino-nucleon
scattering cross section is reduced. Here, we calculate this reduction
explicitly in the perturbative limit $\Gamma_\sigma\ll T$.  By virtue
of an exact sum rule of the spin-density structure function, we
express the modified cross section in terms of the matrix element for
neutrino-nucleon scattering in the presence of a spin-dependent
nuclear potential.  This representation allows for a direct comparison
with and confirmation of Sawyer's related perturbative result.  In a
supernova core with a typical temperature $T=10\,\rm MeV$, the
perturbative limit is relevant for densities $\rho\alt10^{13}\,\rm
g\,cm^{-3}$ and thus applies around the neutrino sphere. There, the
cross-section reduction is of order a few percent and thus not large;
however, a new mode of energy transfer between neutrinos and nucleons
is enabled which may be important for neutrino spectra formation.  We
derive an analytic perturbative expression for the rate of energy
transfer.
\end{abstract}
\vskip2.2pc]


\section{Introduction}

Neutrino scattering rates in a medium differ from those taking place
in vacuum. It is well known that spatial correlations between the
locations or spins of the target particles can reduce or enhance the
average effective scattering cross section. For example, the
anticorrelations caused by the Pauli exclusion principle are
straigthforward to include.  Even in a nondegenerate medium,
correlations are induced by forces between the targets such as the
Coulomb force which thereby causes electromagnetic screening effects
\cite{Sitenko}.  Similarly, in a nuclear medium the spin-dependent
nature of the nucleon-nucleon interaction may cause nonnegligible
``pairings'' of the nucleon spins and thus a reduction of the
axial-current neutrino-nucleon scattering rate \cite{SawyerOld}.

We presently study a less familiar cross-section modification which is
caused by temporal fluctuations rather than spatial correlations. The
main idea is that the neutrino scattering process takes a certain
amount of time.  If the energy transfer is $\omega$, the weak probe
cannot \hbox{``resolve''} those temporal changes of the target
configuration which take place on a time scale faster than about
$1/\omega$. For example, the target nucleon spin may flip ``during''
the neutrino-nucleon collision and thus ``cancel itself.'' In
linear-response theory, this effect is formally described by the
frequency dependence of the nucleon dynamical spin-density structure
function, which in the relevant limit amounts to the Fourier transform
of the autocorrelation function of a single nucleon spin.  In the
absence of interactions the nuclear spin and thus its autocorrelation
function is constant. In the presence of a spin-dependent random force
the initial spin direction is forgotten, causing the spin
autocorrelation function to decrease to zero for large times. Loosely
speaking, then, for small $\omega$ (large ``duration'' of the
collision) the weak probe sees a reduced average target spin and thus
scatters\break less efficiently.

A complete treatment should simultaneously include spin fluctuations
and spin-spin correlations, and presumably spin waves as well. All
these effects are embedded in the dynamical spin-density structure
function, which in general has multiple isospin components. However,
in contrast to spin-spin correlations, spin fluctuations occur even
when there is only one nucleon---provided that its spin is jiggled
around. This is a multiple-scattering effect, not a many-body
phenomenon.  In certain circumstances a pion condensate
\cite{pioncondensate} or the walls in the nuclear bubble phase
\cite{bubblephase} may be the dominant cause for nucleon spin
fluctuations so that it is not entirely academic to study spin
fluctuations independently from spin-spin correlations.

Collision-induced changes of particle velocities or spins cause the
bremsstrahlung emission of photons, neutrino pairs, or axions.
According to the Landau-Pomeranchuk-Migdal (LPM)
effect\cite{Landau,Knoll} the low-energy part of the radiation
spectrum is suppressed if multiple interactions destroy the temporal
coherence of the source. The spin-fluctuation effects studied here are
analogous, except that it is the neutrino scattering rate that is
being reduced.  While the LPM effect is usually discussed for
vector-current processes and thus for velocity fluctuations, in the
case of axial-current processes in nonrelativistic nuclear matter the
spin fluctuations are more significant. We note that temporal
fluctuations do not occur for a conserved quantity such as the charge
of a particle. The vectorial nucleon quantity that does fluctuate due
to collisions is the velocity, which in the nonrelativistic limit is
small. Therefore, in this limit multiple-scattering effects are not
important for vector-current neutrino interactions~\cite{RSb}.  Still,
because in vacuum the nonrelativistic neutral-current neutrino-nucleon
scattering cross section is $\sigma=(C_V^2+3C_A^2) G_{\rm F}^2
E_\nu^2/4\pi$, any modification of the axial-current part strongly
affects the total rate.

The importance of multiple scattering is quantified by the spin
fluctuation rate $\Gamma_\sigma$ which roughly represents the inverse
of the time required for the nucleon to forget its initial spin
orientation. This effect is important if $\Gamma_\sigma$ is of order
the typical energy of the weakly interacting particles which scatter
off, or are emitted from, the medium \cite{RSa}, i.e.\ for
$\Gamma_\sigma\agt T$. One can easily estimate (Eq.~\ref{E4A} below)
that in a supernova (SN) core with a temperature of order $10\,\rm
MeV$ this ``high-density case'' obtains for $\rho\agt 10^{13}\,\rm
g\,cm^{-3}$. Because densities as large as $10^{15}\,\rm g\,cm^{-3}$
are encountered in a SN core, quantities like the neutrino opacity or
the axion emissivity are impossible to calculate in a purely
perturbative way which is based on the assumption that average
scattering or emission rates are the incoherent sum of
single-scattering events. Interaction rates calculated in the ``vacuum
limit'' are fundamentally flawed for the conditions of a SN core.

\begin{figure}[ht]
\centering\leavevmode
\epsfxsize=2.6in
\epsfbox{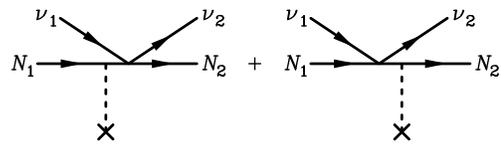}
\smallskip
\caption{Neutrino-nucleon scattering in the presence of an external
spin-dependent potential for the nucleons. The potential can arise
from bystander nucleons, a pion condensate, the walls in the nuclear
bubble phase, or some abstract external force.}
\label{Fig1}
\end{figure}

To extract meaningful estimates for weak interaction rates one must
take recourse to the more general principles of linear-response
theory.  With our collaborators we have begun to develop this
perspective in a series of papers \cite{RSb,KJR,KJRS,Sigl}. We have
argued that the neutrino opacities or axion emissivities can be
estimated by virtue of a phenomenological ansatz for the spin-density
structure function which incorporates certain limiting cases, notably
the low-density one, and which satisfies certain general principles,
in particular a sum rule which can be derived independently of
perturbation theory.  Specifically, we estimated the spin-density
structure function for large energy transfers $\omega$ using a
quasi-bremsstrahlung amplitude (Fig.~\ref{Fig1}).  For small $\omega$,
the corresponding neutrino scattering rate diverges as $1/\omega^2$
due to the intermediate nucleon going on-shell.  Because the true
differential scattering cross section must be finite for all $\omega$,
and motivated by considerations of multiple scattering, we advocated
replacing $1/\omega^2$ by a Lorentzian $1/(\omega^2 + \Gamma^2/4)$
where $\Gamma$ is of order $\Gamma_\sigma$, but is adjusted so that
the structure function obeys the sum rule.

\begin{figure}[ht]
\centering\leavevmode
\epsfxsize=2.6in
\epsfbox{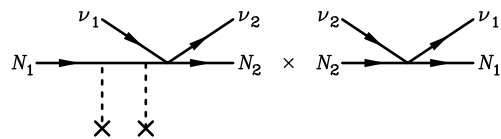}
\smallskip
\caption{One diagram representing the interference between zeroth and
second order scattering amplitudes that leads to a wave-function
renormalization of the incident nucleon for $\nu N$ scattering.}
\label{Fig2}
\end{figure}

Meanwhile, Sawyer \cite{Sawyer} has published an explicit treatment of
the cross-section reduction based on more traditional perturbative
techniques. In addition to the quasi-bremsstrahlung graphs of
Fig.~\ref{Fig1} he includes wave-function and vertex renormalizations
to elastic scattering. The leading correction in the nucleon
scattering potential, $V$, is the interference between zeroth and
second order amplitudes, an example of which is shown in
Fig.~\ref{Fig2}. These terms diverge, behaving as $\delta(\omega)$ in
the absence of nuclear recoil.  However, Sawyer points out, the sum of
all order $|V|^2$ contributions yields a total $\nu N$ cross section
which is finite, but reduced from the vacuum value.

Motivated by Sawyer's work, we show how the divergence of the
quasi-bremsstrahlung process represented by Fig.~\ref{Fig1} can be
rigorously controlled by virtue of our exact sum rule of the
spin-density structure function without assuming any specific
modification of its form, Lorentzian or otherwise, and without
calculating the renormalization terms explicitly.  However, even
though our Lorentzian ansatz is not needed to obtain the perturbative
cross-section reduction effect, it nevertheless yields the correct
limiting value because this ansatz incorporates the sum rule
explicitly.  Put differently, we may implement the sum rule by an
explicit ansatz for the low-$\omega$ behavior of the spin-density
structure function, or we may use the sum rule in an abstract
sense. Either way, in the perturbative limit the final result agrees
with the one found by Sawyer \cite{Sawyer} even though the path of
derivation is entirely different.  Our novel technique has the added
benefit that after the nature of the perturbative region has been
understood, the nonperturbative regime may still be studied using our
proposed Lorentzian modification or some other related ansatz.

In Sect.~II we use the structure-function formalism to derive the
perturbative limit of the average axial-current neutrino-nucleon
scattering cross section. In Sect.~III we consider nucleons
interacting with an external classical potential. In this generic
example the relationship between the perturbative bremsstrahlung
matrix element (Fig.~\ref{Fig1}) and the cross-section reduction
becomes particularly transparent and allows for a direct comparison
with Sawyer's \cite{Sawyer} result.

In a dilute medium where the perturbative approximation is justified,
the most important practical consequence of nucleon spin fluctuations
may not be the mild cross-section reduction, but a new mode of energy
transfer between neutrinos and the nuclear medium\cite{KJRS}. This
energy exchange is enabled by the nontrivial frequency dependence of
the spin-density structure function and thus is specific to spin
fluctuations; spin-spin correlations do not contribute. Indeed it is
plainly visible from the bremsstrahlung nature of the underlying
matrix element (Fig.~\ref{Fig1}) that neutrinos can transfer energy to
nucleons above and beyond the standard nucleon\break recoil effect.

Complementing the numerical expression of Ref.~\cite{KJRS} we derive
in Sect.~IV an analytic expression for the average energy transfer per
collision.  This perturbative result is relevant for conditions around
the neutrino sphere in a SN and thus for the formation of neutrino
spectra.  Sect.~V is given over to discussion and a summary.


\section{Average Neutrino Scattering Rate}

\subsection{Low-Density Limit}

The impact of nucleon spin fluctuations on neutrino scattering rates
is most easily understood in the long-wavelength limit (see Ref.\
\cite{RSb} for a discussion) which has been employed in virtually all
previous papers dealing with neutrino opacities, or neutrino pair and
axion emissivities, in SN cores or old neutron stars.  In this limit,
the momentum transfer between neutrinos and nucleons is neglected. The
axial-current scattering cross section may then be written~as
\begin{equation}
\frac{d\sigma_A}{d\varepsilon_2}=\frac{3C_A^2G_{\rm F}^2}{4\pi}\,
\frac{\varepsilon_2^2\,S(\varepsilon_1-\varepsilon_2)}{2\pi},
\label{E1}
\end{equation}
where $\varepsilon_1$ and $\varepsilon_2$ are the initial- and
final-state neutrino energies, $G_{\rm F}$ is the Fermi constant, and
the neutral-current axial weak coupling constant in a dilute medium is
$C_A \approx +1.37$ and $-1.15$ for protons and neutrons, respectively
\cite{RSb}.

For simplicity we focus on an isotropic, nonrelativistic,
nondegenerate medium of baryon density $n_B$, temperature $T$, and a
single species of nucleons. In this case the function $S(\omega)$ is
the dynamical spin-density structure function in the ${\bf k}\to 0$
limit \cite{RSb,IP}
\begin{equation}
S(\omega)=
\frac{4}{3n_B}\int_{-\infty}^{+\infty} dt\,
e^{i\omega t}\langle\hbox{\boldmath$\sigma$}(t)\cdot
\hbox{\boldmath$\sigma$}(0)\rangle,
\label{E2}
\end{equation}
where $\langle\hbox{\boldmath$\sigma$}(t)\cdot
\hbox{\boldmath$\sigma$}(0)\rangle$ is the autocorrelation
function for the nucleon spin operator
$\hbox{\boldmath$\sigma$}(t)=\int d^3{\bf x}\
\hbox{\boldmath$\sigma$}(x)$ at time $t$. Here
$\hbox{\boldmath$\sigma$}(x)\equiv\frac{1}{2}\psi^\dagger(x)
\hbox{\boldmath$\tau$}\psi(x)$, $\psi(x)$ is the nucleon field
(a Pauli two-spinor) and $\hbox{\boldmath$\tau$}$ is a vector of Pauli
matrices. The expectation value $\langle\ldots\rangle$ is taken over a
thermal ensemble so that detailed balance
$S(\omega)=S(-\omega)\,e^{\omega/T}$ is satisfied. We note that our
definition of energy transfer is positive for energy given to the
medium.

In order to derive an average scattering cross section we consider
nondegenerate thermal neutrinos which we take to follow a
Maxwell-Boltzmann distribution; the difference to a Fermi-Dirac
distribution is inessential for the present discussion. Therefore, we
consider the quantity
\begin{equation}
\langle \sigma_A\rangle
\equiv \frac{3C_A^2G_{\rm F}^2}{4\pi}\,
\frac{\int d^3{\bf k_1}\,
e^{-\varepsilon_1/T} \int_0^\infty d\varepsilon_2\,
\varepsilon_2^2\,S(\varepsilon_1-\varepsilon_2)}
{2\pi \int d^3{\bf k_1}\,
e^{-\varepsilon_1/T}}\,.
\label{E3}
\end{equation}
With the dimensionless energy transfer
$x\equiv (\varepsilon_1-\varepsilon_2)/T$ and after one explicit
integration one finds \cite{RSb}
\begin{equation}
\langle \sigma_A\rangle
= \sigma_T \int_0^\infty \frac{dx}{2\pi}\,\widetilde S(x)\,
(2+x+{\textstyle{1\over6}}x^2)\,e^{-x}.
\label{E4}
\end{equation}
Here, $\sigma_T\equiv \frac{9}{\pi}\,C_A^2G_{\rm F}^2\,T^2$ while
$\widetilde S(x)\equiv T\,S(x T)$ is the dimensionless structure
function. In vacuum the nucleon spins do not evolve, yielding
a constant autocorrelation function and thus 
$\widetilde S(x)=2\pi\delta(x)$. Then 
$\langle \sigma_A\rangle=\sigma_T$ where
$\int_0^\infty dx\, \delta(x)=\frac{1}{2}$ has been used.

\begin{figure}[ht]
\centering\leavevmode
\epsfxsize=2.4in
\epsfbox{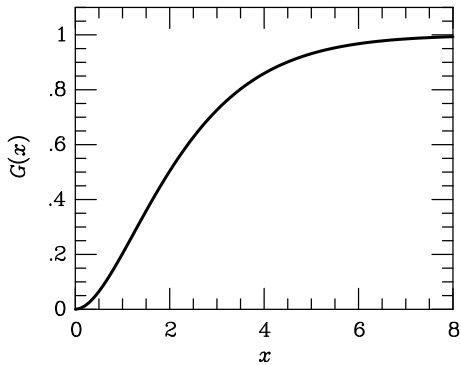}
\smallskip
\caption[...]{Function $G(x)$ as defined in Eq.~(\ref{E8}).}
\label{Fig3}
\end{figure}

For reasons that will soon become apparent, we concentrate not on a
direct calculation of the average cross-section at finite density
$\langle\sigma_A\rangle$, but rather on its deviation from the vacuum
cross section $\delta\langle \sigma_A \rangle \equiv
\langle \sigma_A \rangle-\sigma_T$ or
\begin{equation}
\frac{\delta\langle \sigma_A \rangle}{\sigma_T}
=-1+\int_0^\infty \frac{dx}{2\pi}\,\widetilde S(x)\,
(2+x+{\textstyle{1\over6}}x^2)\,e^{-x}.
\label{E5}
\end{equation}
The crucial step is to express the r.h.s.\ as a common integral over
$\widetilde S(x)$. To this end we use the normalization 
$\int_{-\infty}^{+\infty}\widetilde S(x)\,dx/2\pi=1$ which obtains if
the spins of different nucleons evolve independently; otherwise an
additional correlation term would appear on the r.h.s.\
\cite{RSb,KJRS,Sigl}, a possibility to be addressed in Sect.~II.B
below. By virtue of detailed balance this sum rule is
\begin{equation}
\int_0^\infty \frac{dx}{2\pi}\,\widetilde S(x)\,(1+e^{-x})=1.
\label{E6}
\end{equation}
Replacing $-1$ in Eq.~(\ref{E5}) by the negative of Eq.~(\ref{E6})
yields
\begin{equation}
\frac{\delta\langle \sigma_A \rangle}{\sigma_T}
=-\int_0^\infty \frac{dx}{2\pi}\,\widetilde S(x)\,G(x),
\label{E7}
\end{equation}
where $G(x)\equiv-[(2+x+\frac{1}{6}\,x^2)\,e^{-x} -(1+e^{-x})]$ or
\begin{equation}
G(x)= 1 - (1+x+{\textstyle{1\over6}}\,x^2)\,e^{-x}.
\label{E8}
\end{equation}
This function is shown in Fig.~\ref{Fig3}. It expands as
$G(x)=\frac{1}{3}\,x^2+{\cal O}(x^3)$ for small $x$, approaches $1$
for large $x$, and is always positive. Because $\widetilde S(x)$ is
also a positive function, we find that the average cross section in
the medium is indeed always suppressed by spin fluctuations.

Thus far $S(\omega)$ has been the nonperturbative but unknown
structure function. However, what can be calculated in the framework
of perturbation theory is an expression $S_{\rm brems}(\omega)$
based on the ``bremsstrahlung'' or ``medium excitation'' amplitude
Fig.~\ref{Fig1}. In our previous works \cite{RSb,RSa}, we showed that
$S_{\rm brems}(\omega)$ diverges for small $\omega$ as $\omega^{-2}$,
a behavior which is generic for all bremsstrahlung processes---for the
electromagnetic case see Jackson \cite{Jackson}.  We may then write
\begin{equation}
S_{\rm brems}(\omega)=\frac{\Gamma_\sigma}{\omega^2}\,s(\omega/T)
\times\cases{e^{\omega/T}&for $\omega<0$,\cr
\noalign{\vskip3pt plus1pt}
1 &for $\omega>0$,\cr}
\label{E10}
\end{equation}
where $s(x)$ is a nonsingular even function with $s(0)=1$. The
quantity $\Gamma_\sigma$, defined as the coefficient of the
$\omega^{-2}$ singularity of $S_{\rm brems}(\omega)$, is physically
interpreted as the spin-fluctuation rate.\footnote{One may consider
the limit of a classical spin vector ${\bf s}(t)$ being kicked by a
random force at a rate $\Gamma_{\rm coll}$. If the spin changes
abruptly by a random amount $\Delta {\bf s}$ in a given collision
(which is thus assumed to be ``hard'') and if subsequent spin
orientations are uncorrelated one finds
$S_{\rm brems}(\omega)=\Gamma_\sigma/(\omega^2+\Gamma_\sigma^2/4)$
with $\Gamma_\sigma=\Gamma_{\rm coll}\,
\langle(\Delta {\bf s})^2\rangle/\langle{\bf s}^2\rangle$
\cite{Knoll,Raffelt}. This justifies identifying $\Gamma_\sigma$ with
an average spin rate of change or a spin-fluctuation rate. In the
classical limit of hard collisions one has $s(x)=1$, while for general
interaction potentials $s(x)$ is more complicated. Quantum corrections
introduce the detailed-balance factor, and cause $s(x)$ to be a
decreasing function for large $x$, as discussed for the case of
electromagnetic bremsstrahlung by Jackson \cite{Jackson}. The same
conclusion is inferred from the f-sum rule for $S(\omega)$
\cite{Sigl}.}

It should now be clear why we have calculated the deviation
$\delta\langle\sigma_A\rangle$ rather than $\langle\sigma_A\rangle$
itself---the $x^2$ behavior of $G(x)$ compensates for the singularity
in $S_{\rm brems}$. We are therefore free to substitute
$S_{\rm brems}(\omega)$ for $S(\omega)$ and are assured of a finite
answer for $\delta\langle\sigma_A\rangle$. Further, if we accept that
in a dilute medium the true $S(\omega)$ is well-represented by
$S_{\rm brems}(\omega)$ for $\omega \gg \Gamma_\sigma$ then
\begin{equation}
\frac{\delta\langle \sigma_A \rangle}{\sigma_T}
=-\int_0^\infty \frac{dx}{2\pi}\,\widetilde S_{\rm brems}(x)\,G(x)
\label{E11}
\end{equation}
is the desired perturbative result.

With our representation Eq.~(\ref{E10}) the cross-section reduction is
to lowest order in $\gamma_\sigma\equiv\Gamma_\sigma/T$
\begin{equation}
\frac{\delta\langle \sigma_A \rangle}{\sigma_T} =
-\frac{\gamma_\sigma}{2\pi} \int_0^\infty
dx\,\frac{G(x)}{x^2}\,s(x).
\label{E12}
\end{equation}
Taking for simplicity the classical limit $s(x)=1$ we find
\begin{equation}
\frac{\delta\langle \sigma_A \rangle}{\sigma_T} =
-\frac{5}{6}\,\frac{\gamma_\sigma}{2\pi}.
\label{E13}
\end{equation}
Once more, these results put in evidence that $\gamma_\sigma$ is the
expansion parameter which defines the perturbative regime.

We may estimate the error due to using $\widetilde S_{\rm brems}(x)$
in Eq.~(\ref{E11}) instead of the full $\widetilde S(x)$ in
Eq.~(\ref{E7}). If the true $\widetilde S(x)$ is given by 
$\widetilde S_{\rm brems}(x)$ to lowest order in $\gamma_\sigma$ so 
that $\widetilde S(x)-\widetilde S_{\rm brems}(x)=
{\cal O}(\gamma_\sigma^2)$ for $x\gg\gamma_\sigma$, then
$\int_0^\infty[\widetilde S(x)-\widetilde S_{\rm brems}(x)] 
\,G(x)\,dx/2\pi={\cal O}(\gamma_\sigma^2)$. 

This implies that the lowest-order cross-section reduction effect
represented by Eq.~(\ref{E12}) will be found by any assumed functional
form $S_{\rm approx}(\omega)$ for the true $S(\omega)$ if $S_{\rm
approx}(\omega)$ agrees with $S_{\rm brems}(\omega)$ to ${\cal
O}(\gamma_\sigma^2)$ for $\omega\gg\Gamma_\sigma$. Any such function
which is normalized can be inserted into Eq.~(\ref{E4}) and will then
yield Eq.~(\ref{E12}) up to an error of 
${\cal O}(\gamma_\sigma^2)$. Further, any such function, even if it is
not normalized, will yield this result when inserted into
Eq.~(\ref{E7}) where the normalization condition has been reshuffled
into the function $G(x)$. To lowest order in $\gamma_\sigma$, the
cross-section reduction effect is independent of the detailed
structure of the true $S(\omega)$ in the neighborhood of $\omega=0$.

\subsection{Spin-Spin Correlations}

A crucial step in the above analysis was use of the sum rule in
Eq.~(\ref{E6}), appropriate for a medium of uncorrelated nucleons.
However, in a real nuclear medium the nucleon spin fluctuations are
typically caused by a spin-dependent interaction among nucleons.
Inevitably, this will cause correlations between different spins so
that the r.h.s.\ of the sum rule Eq.~(\ref{E6}) is
$1+C(\gamma_\sigma)$ where in a dilute medium
$|C(\gamma_\sigma)|\ll 1$. It follows that $G(x)$
receives an additional contribution $-C(\gamma_\sigma)(1+e^{-x})$ and
$\delta \langle\sigma_A\rangle/\sigma_T$ one of order
$C(\gamma_\sigma)$.  Here $\delta \langle\sigma_A\rangle/\sigma_T$ is
to be calculated with the full $\widetilde S(x)$ not 
$\widetilde S_{\rm brems}(x)$. If $C(\gamma_\sigma)$ is of order 
$\gamma_\sigma^n$, then the correction to the cross-section shift 
from considering spin-correlations is also of order $\gamma_\sigma^n$.

We may next use the above estimate of the error incurred by using 
$\widetilde S_{\rm brems}(x)$ rather than the true 
$\widetilde S(x)$. Then, if $C(\gamma_\sigma) \propto 
\gamma_\sigma^n$, with $n>1$ the cross-section deviation 
calculated from $\widetilde S_{\rm brems}(x)$ in Eq.~(\ref{E11}) is 
to lowest order independent of spatial spin-spin correlations.

For example, if the nucleon-nucleon interaction potential is written
as in Ref.~\cite{Sigl}, and if the correlation length scales as
$\gamma_\sigma$, then we expect $C(\gamma_\sigma)$ to be of
order $\gamma_\sigma^3$, in which case the low-density limit of the
cross-section change is well described by Eq.~(\ref{E11}).

\subsection{Comparison with the High-Density Behavior}

We next compare the low-density limit thus derived with our more
general previous expression. In a dense medium
(\hbox{$\gamma_\sigma\agt1$}) the detailed structure of $S(\omega)$
for low energy transfers matters. In the past we have advocated a
Lorentzian form
\begin{equation}
S_{\rm approx}(\omega)=\frac{\Gamma_\sigma}{\omega^2+\Gamma^2/4}\,
s(\omega/T)
\times\cases{e^{\omega/T}&for $\omega<0$,\cr
\noalign{\vskip3pt plus1pt}
1&for $\omega>0$,\cr}
\label{E15}
\end{equation}
where for a given $\Gamma_\sigma$ one chooses $\Gamma$ such that
$S_{\rm approx}(\omega)$ is normalized. This ansatz is motivated by a
heuristic argument \cite{RSa} and by the classical limit which obtains
for $\omega\ll T$ \cite{Knoll,Raffelt}. Equation~(\ref{E15}) naturally
approaches the appropriate limit for low densities.

\begin{figure}[ht]
\centering\leavevmode
\epsfxsize=2.4in
\epsfbox{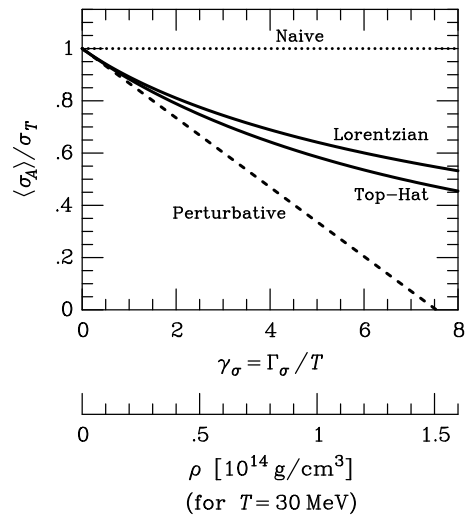}
\medskip
\caption[...]{Average axial-current neutrino-nucleon scattering cross
section as a function of the spin-fluctuation rate in the classical
limit where $s(x)=1$. The dotted line refers to a naive calculation
where nucleon spin fluctuations are ignored entirely. The dashed line
refers to the low-density perturbative expansion which agrees with
Sawyer's \cite{Sawyer} related result if one expresses the
spin-fluctuation rate $\Gamma_\sigma$ in terms of the assumed nucleon
interaction potential (Sect.~III.A).  The solid line marked
``Lorentzian'' arises from the ansatz Eq.~(\ref{E15}) while the one
marked ``Top-Hat'' arises from Eq.~(\ref{E15x}).  The second
horizontal axis expresses $\gamma_\sigma$ in terms of the physical
density $\rho$ for $T=30\,\rm MeV$, using the OPE interaction
potential according to Eq.~(\ref{E4A}). The correspondence between
$\gamma_\sigma$ and $\rho$ is meaningful only in the perturbative
regime where $\gamma_\sigma<\hbox{a few}$.}
\label{Fig4}
\end{figure}

In Fig.~\ref{Fig4} we show $\langle\sigma_A\rangle/\sigma_T$ for
$s(x)=1$ as a function of $\gamma_\sigma$. The dotted line marks the
``naive'' constant cross section which obtains when spin fluctuations
are ignored entirely. The dashed line represents the perturbative
result according to Eq.~(\ref{E13}); for $\gamma_\sigma\agt7.5$ it
yields complete nonsense (a negative scattering cross section). The
solid line marked ``Lorentzian'' was obtained with the above ansatz
for $S_{\rm approx}(\omega)$. The dashed line is its tangent at the
point $\gamma_\sigma=0$ so that indeed the Lorentzian ansatz yields
the same perturbative limit as the direct calculation in Sect.~II.A
where the sum rule was implemented in an abstract sense rather than by
a specific ansatz for the low-$\omega$ behavior of $S(\omega)$. The
Lorentzian ansatz yields a plausible intermediate result between the
naive and lowest-order perturbative results.

The overall shape of the Lorentzian line in Fig.~\ref{Fig4} is
determined by the bremsstrahlung wings of $S_{\rm approx}(\omega)$
together with the sum rule. In order to test how sensitive it is to
the assumed low-$\omega$ shape we have considered a second ansatz of
the form
\begin{equation}
S_{\rm approx}(\omega)=\Gamma_\sigma\times
\cases{\omega_0^{-2}&for $0\leq\omega\leq\omega_0$,\cr
\noalign{\vskip3pt plus1pt}
\omega^{-2}&for $\omega_0<\omega$.\cr}
\label{E15x}
\end{equation}
Of course, for $\omega<0$ we have the detailed-balance factor
$e^{\omega/T}$ as in Eq.~(\ref{E15}), and in general there is a
function $s(\omega/T)$ which we take to be equal to 1 for the purpose
of illustration. For a given choice of $\Gamma_\sigma$ the frequency
$\omega_0$ is determined such that $S_{\rm approx}(\omega)$ fulfills
the sum rule. The cross-section reduction derived from this
``top-hat'' ansatz is shown in Fig.~\ref{Fig4}. It has a common
tangent at $\gamma_\sigma=0$ with the dashed line and the Lorentzian
curve, again confirming that in the perturbative limit the detailed
low-$\omega$ shape of $S(\omega)$ does not matter. For large
$\gamma_\sigma$ the deviation from the Lorentzian curve is relatively
small.  Therefore, it appears that even in the nonperturbative regime
the cross-section reduction is dominated by the bremsstrahlung
calculation in conjunction with the sum-rule, and not by fine points
of the low-$\omega$ shape of~$S(\omega)$.

In this discussion we have used the classical brems\-strah\-lung limit
of hard collisions where $s(x)=1$. Quantum corrections alone require
that $s(x)$ is a decreasing function of $x$ for large $x$, and the
same conclusion is reached on the basis of Sigl's f-sum rule
\cite{Sigl}.  Further, the detailed large-$x$ behavior depends on the
short-distance behavior of the assumed $N\!N$ interaction
potential. It is evident from Eq.~(\ref{E12}) that the detailed
functional form of $s(x)$ will determine the slope of the curves in
Fig.~\ref{Fig4} at $\gamma_\sigma=0$. However, any 
change in slope is
independent of the manner by which $\widetilde S(x)$ has been adjusted
to satisfy the sum rule.  Thus, the three curves remain tangent to
each other, although their slope will in general be different from the
value $-\frac{5}{6}\frac{1}{2\pi}$ determined for $s(x)=1$.

To express the spin-fluctuation rate in terms of the physical
density and temperature we need to assume a specific model for the
cause of the spin fluctuations. Taking nucleon-nucleon
interactions modelled by a one-pion exchange (OPE) potential as in
our previous papers one finds for a single species of nucleons
\begin{equation}
\Gamma_{\sigma, {\rm OPE}} = 4\sqrt\pi\,\alpha_\pi^2
\frac{n_B T^{1/2}}{m_N^{5/2}}
= 8.6\,{\rm MeV}\,\rho_{13}\,T_{10}^{1/2},
\label{E4A}
\end{equation}
where $\alpha_\pi\equiv(f 2m_N/m_\pi)^2/4\pi\approx15$ with $f\approx
1.0$ is the pion fine structure constant,
$\rho_{13}\equiv\rho/10^{13}\,\rm g\,cm^{-3}$,
$T_{10}\equiv T/10\,{\rm MeV}$, and $m_N=940\,\rm MeV$ is the nucleon
mass. The pion mass has been neglected.  Taking $T\approx 10\,\rm MeV$
as a typical value for SN conditions,  one
concludes that the dividing line between high and low density is
roughly given by $10^{13}\,\rm g\,cm^{-3}$ or 3\% nuclear density.

We stress that $\Gamma_{\sigma,\rm OPE}$ is in itself a perturbative
result and thus will be a reasonable representation of the true
$\Gamma_\sigma$ only if $\Gamma_{\sigma,\rm OPE}/T\alt 1$. Therefore,
in the high-density regime Eq.~(\ref{E4A}) cannot be used to translate
an assumed spin-fluctuation rate into a corresponding physical
density. We have previously argued that the true $\gamma_\sigma$ in a
nuclear medium never exceeds a few \cite{KJRS,Sigl}.


\section{Matrix-Element Representation}

\subsection{Perturbative Cross-Section Reduction}

The perturbative structure function $S_{\rm brems}(\omega)$ is
calculated from the quasi-bremsstrahlung process shown in
Fig.~\ref{Fig1} so that one may represent the cross-section reduction
$\delta\langle\sigma_A\rangle$ directly in terms of its matrix element
${\cal M}$. The translation is most easily achieved by considering the
differential scattering cross section. Denoting the four-momentum
of the in- and outgoing nucleon with
$(E_1,{\bf p}_1)$ and $(E_2,{\bf p}_2)$, respectively, we find
\eject

\begin{eqnarray}
\frac{d\sigma_A}{d\varepsilon_2}&=&
\frac{n_c}{n_B}\,\varepsilon_2^2
\int \frac{d\Omega_2}{(2\pi)^3}
\frac{d^3{\bf p}_1}{(2\pi)^3}\,\frac{d^3{\bf p}_2}{(2\pi)^3}\,
\frac{d^3{\bf k}}{(2\pi)^3}\,
f_1\nonumber\\ \noalign{\vskip4pt plus1pt}
&\times&\sum_{\rm spins}
\frac{\left\langle|{\cal M}|^2\right\rangle}
{2\varepsilon_1 2\varepsilon_2 2E_1 2E_2}
\,(2\pi)^3\,\delta^3({\bf p}_1+{\bf k}-{\bf p}_2)\nonumber\\
\noalign{\vskip4pt plus1pt} &\times&2\pi\,\delta(E_1+\varepsilon_1-
E_2-\varepsilon_2)\,.
\label{E20}
\end{eqnarray}
Here, $n_c$ is the number density of classical scattering centers,
$f_1$ is the occupation number of the initial-state nucleon, ${\bf k}$
is the momentum absorbed by the external potential, Pauli blocking
factors are ignored for all particles because of the assumed
nondegeneracy, and the neutrinos have been ignored in the momentum
$\delta$ function because of the long-wavelength approximation. The
expectation value $\langle|{\cal M}|^2\rangle$ is understood to
include the averaging of classical ensemble variables on which the
external potential might depend. The perturbative structure function
$S_{\rm brems}(\omega)$ is obtained by comparing Eq.~(\ref{E20}) with
Eq.~(\ref{E1}).

In order to derive the matrix element we use the axial part of the
weak interaction Hamiltonian and an external classical potential for
the nucleon spins. The most 
general form for the potential in Fourier space is~\cite{BD}
\begin{eqnarray}
  V({\bf k},\hbox{\boldmath $\sigma$},{\bf s})&=&
  U_0(k)+U_{\rm S}(k)\,\hbox{\boldmath $\sigma$}
  \cdot{\bf s}\nonumber\\
  &+&U_{\rm T}(k)\,(3\hbox{\boldmath $\sigma$}\cdot\hat{\bf k}\,
  {\bf s}\cdot\hat{\bf k}-\hbox{\boldmath $\sigma$}\cdot{\bf s})
  \,,\label{Vint}
\end{eqnarray}
where $k=|{\bf k}|$, $\hat{\bf k}={\bf k}/k$, ${\bf s}$ is a classical
spin vector of length 1 associated with the external potential, and
\hbox{\boldmath$\sigma$} is the nucleon spin operator.
Here, $U_0$ is a spin-independent potential while $U_{\rm S}$ and
$U_{\rm T}$ represent a spin-dependent scalar and tensor force, respectively.
After some algebra one finds
\begin{eqnarray}
  \sum_{\rm spins}\frac{\left\langle|{\cal M}|^2\right\rangle}
  {2\varepsilon_1 2\varepsilon_1 2E_1 2E_2}
  &=&\frac{G_{\rm F}^2C_A^2}{2\omega^2}
  \Bigl[\bigl|U_{\rm S}(k)\bigr|^2
  \bigl(1-{\textstyle{1\over3}}c_{12}\bigr) \nonumber\\
  &&\kern-3em-\,2{\rm Re}\bigl[U_{\rm T}(k)U_{\rm S}^*(k)\bigr]\,
  \bigl(c_1 c_2- {\textstyle{1\over3}}c_{12}\bigr)\nonumber\\
  \noalign{\vskip3pt}
  &&\kern-3em+\,\bigl|U_{\rm T}(k)\bigr|^2
  \bigl(2-c_1 c_2-{\textstyle{1\over3}}c_{12}\bigr)\Bigr],\label{matr}
\end{eqnarray}
where $c_i$ ($i=1$ or 2) is the cosine of the angle between the
direction of neutrino $i$ relative to ${\bf k}$, while $c_{12}$
refers to the angle between the two neutrinos. We have averaged over
the external spin directions ${\bf s}$ with an assumed isotropic
distribution.

The interaction $U_0(k)$ does not contribute because it leaves the
nucleon spins unchanged. This leaves us with the scalar and tensor
force $U_{\rm S}(k)$ and $U_{\rm T}(k)$, respectively. If the
classical scatterers are substituted by the nucleons themselves only
the tensor term survives because the scalar term conserves the total
spin of two colliding nucleons and thus does not cause spin
fluctuations~\cite{Sigl}.

Expression (\ref{matr}) reveals explicitly the $\omega^{-2}$
divergence of Eq.~(\ref{E20}) which thus cannot be integrated to
yield a total cross section.  However, following the steps of
Sect.~II.A we can derive a convergent expression for the
medium-induced {\it change\/} of $\sigma_1$ which denotes the total
axial-current scattering cross section for a fixed initial-state
energy $\varepsilon_1$. In the structure-function language it is the
$d\varepsilon_2$ integral of Eq.~(\ref{E1}) or equivalently
\begin{equation}
\sigma_1=\frac{3C_A^2G_{\rm F}^2}{4\pi}\,
\int_{-\infty}^{+\infty} \frac{d\omega}{2\pi}\,S(\omega)\,
(\varepsilon_1-\omega)^2\,\Theta(\varepsilon_1-\omega).
\label{E22}
\end{equation}
In vacuum $\sigma_{1,\rm vac}= (3C_A^2G_{\rm F}^2/4\pi)\,
\varepsilon_1^2$ so that the medium-induced change
$\delta\sigma_1\equiv \sigma_{1,\rm med}-\sigma_{1,\rm vac}$ is
\begin{equation}
\frac{\delta\sigma_1}{\sigma_{1,\rm vac}}
=-1+\int_{-\infty}^{+\infty} \frac{d\omega}{2\pi}\,S(\omega)\,
\frac{(\varepsilon_1-\omega)^2\,\Theta(\varepsilon_1-\omega)}
{\varepsilon_1^2}\,.
\label{E23}
\end{equation}
Then we may proceed as before and replace $-1$ by an integral over the
structure function by virtue of its normalization so that\footnote{In
this form one can easily see that for small $\varepsilon_1$ the cross
section actually increases. For example, $\varepsilon_1=0$ leads to a
vanishing vacuum cross section while in the medium it is $(3 C_A^2
G_{\rm F}^2/4\pi)\,\int_{-\infty}^0 d\omega\,\omega^2\,
S(\omega)/2\pi$ or by detailed balance $(3 C_A^2 G_{\rm
F}^2/4\pi)\,\int_0^{\infty} d\omega\,\omega^2\,
S(\omega)\,e^{-\omega/T}/2\pi$.}
\begin{equation}
\frac{\delta\sigma_1}{\sigma_{1,\rm vac}}=
\int_{-\infty}^{+\infty} \frac{d\omega}{2\pi}\,S(\omega)\,\left[
\frac{(\varepsilon_1-\omega)^2\, \Theta(\varepsilon_1-\omega)}
{\varepsilon_1^2}-1\right].
\label{E24}
\end{equation}
As before, the integrand varies effectively as $S(\omega)\omega^2$ for
small $\omega$ because the term linear in $\omega$ switches sign
at the origin. Therefore, to lowest order we may substitute
$S(\omega)\to S_{\rm brems}(\omega)$, provided we interpret
the remaining integral by its principal part.

$S_{\rm brems}(\omega)$ is obtained by comparing Eq.~(\ref{E1}) with
Eq.~(\ref{E20}) and using Eq.~(\ref{matr}). After performing the
$d\omega$, $d{\bf p}_2$, and $d\Omega_2$ integrations we arrive at
\begin{eqnarray}
\frac{\delta\sigma_1}{\sigma_{\rm 1,vac}}&=&\frac{2}{3}\,
\frac{n_c}{n_B}\int \frac{d^3{\bf p}_1}{(2\pi)^3}\,
\frac{d^3{\bf k}}{(2\pi)^3}\,f_1\nonumber \\
\noalign{\vskip4pt plus1pt}
&&\kern-2em\times
\frac{\bigl|U_{\rm S}(k)\bigr|^2+2\bigl|U_{\rm T}(k)\bigr|^2}
{\omega^2}\,
\left[\frac{(\varepsilon_1-\omega)^2\,
\Theta(\varepsilon_1-\omega)}{\varepsilon_1^2}-1\right],
\nonumber\\
\label{E26}
\end{eqnarray}
where the energy transfer is
$\omega=-(2{\bf p}_1\cdot{\bf k}+{\bf k}^2)/2m_N$.

\subsection{Comparison with Sawyer's Result}

As mentioned in the introduction, Sawyer \cite{Sawyer} has discussed a
cross-section reduction due to the interaction of the target nucleons
with bystander particles. He does not provide an immediate physical
interpretation of his calculation, but we believe that in essence he
has studied the same effect that is the topic of our paper, namely the
scattering version of the Landau-Pomeranchuk-Migdal effect.  However,
his formal approach is quite different from ours.

The optical theorem implies that calculating the total neutrino
scattering cross section amounts to a calculation of the imaginary
part of the neutrino forward-scattering amplitude $f_0$ on
nucleons. Sawyer uses analyticity constraints\cite{analyt} for $f_0$
to recognize that the total cross section should be finite order by
order in a perturbative expansion in powers of the nucleon interaction
potential. Further, he observes that this result holds even though
individual contributions to $f_0$ have infrared singularities from
on-shell intermediate states. One type of $V^2$ contribution to $f_0$
comes from interference between the zeroth and second order scattering
amplitudes. Fig.~\ref{Fig2} shows such a contribution which may be
interpreted as a wave-function renormalization of the incoming
nucleon. Similar terms would renormalize the outgoing nucleon wave
function, or provide a vertex correction. The other type of $V^2$
contribution to ${\rm Im}(f_0)$ is given by a phase-space integral
over the square of the amplitudes shown in Fig.~\ref{Fig1}. These
terms correspond to the quasi-bremsstrahlung inelastic scattering
process, and also diverge as discussed above. Sawyer's main point is
that the divergences in these two types of terms must sum to a finite
result.

Sawyer \cite{Sawyer} has worked out several examples which illustrate
this approach.  Specifically, the cross-section reduction represented
by his Eq.~(10) is very similar to our Eq.~(\ref{E26}). However, our
Eq.~(\ref{E26}) has not yet been averaged over initial state neutrino
energies, Sawyer has used bystander nucleons to provide the potential
so that his expression for the energy transfer takes account of the
bystander recoil, and he has studied $\nu n\to p e^-$ scattering
rather than $\nu N\to N \nu$ so that the proton-neutron mass
difference appears.  Further, he has used a scalar potential which is
explicitly isospin dependent so that the role of our spin fluctuations
is played by isospin fluctuations in his case.

The main point of agreement is the structure of the term in square
brackets in Eq.~(\ref{E26}). Both Sawyer's Eq.~(10) and our
Eq.~(\ref{E26}) diverge if one considers the first term in square
brackets independently from the $-1$. In our derivation, the $-1$
effectively represents the sum rule of the nonperturbative $S(\omega)$
for which we have substituted $S_{\rm brems}(\omega)$ after the two
terms have been combined. In Sawyer's approach, the $-1$ corresponds
to the wave-function renormalization of the elastic scattering rate.
Our interpretation of the agreement between these results is as
follows.

In effect, Sawyer has calculated the perturbative approximation
$S^{(2)}(\omega)$ to second order in $V$. Recall that the zeroth-order
approximation is $S^{(0)}(\omega)=2\pi\delta(\omega)$ because Sawyer
also uses the long-wavelength limit where nucleon recoil effects are
ignored. In this limit, any nonvanishing power of $S^{(2)}(\omega)$ at
$\omega\not=0$ must arise from the quasi-bremsstrahlung amplitudes of
Fig.~\ref{Fig1} so that inevitably $S^{(2)}(\omega)=S_{\rm
brems}(\omega)$ for $\omega\not=0$. Sawyer's renormalization terms
modify only the elastic channel $\omega=0$ so that his complete result
amounts to $S^{(2)}(\omega)=S_{\rm brems}(\omega)-A\delta(\omega)$
where $A$ is an infinite integral expression. Of course,
$S^{(2)}(\omega)$ is highly singular and thus unphysical at $\omega=0$
in the sense that in the neighborhood of $\omega=0$ it does not
provide a representation of the differential scattering cross section
Eq.~(\ref{E1}).  However, $S^{(2)}(\omega)$ is legitimate as an
integral kernel to calculate the total cross section
Eq.~(\ref{E4}). The agreement between Sawyer's and our results shows
that the second-order perturbative calculation yields an expression
for $A$ such that $S^{(2)}(\omega)$ fulfills our sum rule.

In essence, then, Sawyer's calculation amounts to showing explicitly
that the renormalization terms not only cancel the low-$\omega$
divergence of $S_{\rm brems}(\omega)$, but indeed cancel it in such a
way that $S^{(2)}(\omega)$ fulfills the sum rule. The renormalization
terms are an explicit second-order manifestation of the information
embodied in our sum rule.  In our derivation, we have shown the sum
rule to be a general nonperturbative property of
$S(\omega)$. Therefore, once we have calculated $S_{\rm
brems}(\omega)$ we can handle its low-$\omega$ divergence either by an
abstract application of the sum rule, or by an explicit ansatz for the
physical behavior of the true $S(\omega)$ near $\omega=0$.  Either way
we do not need to calculate the renormalziation terms explicitly.

Although Sawyer's and our approaches are equivalent in the low-density
limit, they are not equivalent when one considers the high-density
case. There, a perturbative expansion makes no sense as higher-order
terms exceed the lower-order ones. However, by making use of the sum
rule, and exploiting the physical insight that $S(\omega)$ should have
a width of order $\Gamma_\sigma$ \cite{Sigl} and possess a hard
bremsstrahlung tail for $\omega \gg \Gamma_\sigma$, we have the basis
for a reasonable model of the high-density regime.

To summarize, our derivation is based on representing interaction
rates by virtue of current correlators which allow for a direct
transition to the classical limit. Therefore, our approach allows for
an intuitive interpretation of the cross-section reduction as a
temporal spin-averaging effect. Moreover, because we know on general
grounds that the sum rule Eq.~(\ref{E6}) must be fulfilled, we do not
need to worry about a calculation of the various infinite second-order
corrections to the elastic scattering rate. In our derivation, the
only required Feynman-graph evaluation is that of the ``medium
excitation term'' of Fig.~\ref{Fig1}.  Finally, our derivation allows
for a clear and physical identification of the dimensionless parameter
$\gamma_\sigma$ which defines the perturbative expansion. Sawyer's
technique, on the other hand, represents a more familiar methodology
if one approaches the problem as a perturbative expansion in powers of
$V$.


\section{Energy Transfer}

As stressed in Ref.~\cite{KJRS}, the most important effect of nucleon
spin fluctuations may be that they allow for a new mode of energy
transfer by the quasi-bremsstrahlung process shown in Fig.~\ref{Fig1}.
The relevant figure of merit is the average energy transfer per
collision $\langle\Delta \varepsilon\rangle_{\rm brems}$ or
\begin{equation}
\frac{\int_0^\infty d\varepsilon_1\varepsilon_1^2
e^{-\varepsilon_1/T_\nu} \int_0^\infty
d\varepsilon_2\varepsilon_2^2(\varepsilon_2-\varepsilon_1)
S(\varepsilon_1-\varepsilon_2)}
{\int_0^\infty d\varepsilon_1\varepsilon_1^2 e^{-\varepsilon_1/T_\nu}
\int_0^\infty d\varepsilon_2\varepsilon_2^2 S(\varepsilon_1-
\varepsilon_2)}.
\label{E30}
\end{equation}
Here, $T_\nu$ is the temperature of the neutrinos which are assumed to
follow a Maxwell-Boltzmann distribution while the nucleons are
characterized by $T$. In Ref.~\cite{KJRS} this expression was
evaluated numerically on the basis of the Lorentzian ansatz for the
structure function.

However, in the dilute-medium limit one can also derive an explicit
expression. We first note that in the numerator and denominator one
can each perform one integration explicitly so that
\begin{equation}
\frac{\langle\Delta \varepsilon\rangle_{\rm brems}}{T}
=\frac{\int_0^\infty dx\,\widetilde S(x)\,x\,F_\beta(x)\,
(e^{-x}-e^{-\beta x})}
{\int_0^\infty dx\,\widetilde S(x)\,F_\beta(x)\,
(e^{-x}+e^{-\beta x})}
\label{E31}
\end{equation}
with
$F_\beta(x)\equiv 1+{\textstyle{1\over2}}\beta
x+{\textstyle{1\over{12}}}\,\beta^2x^2$ and $\beta\equiv T/T_\nu$.

In the dilute-medium limit we may use to lowest order
$\widetilde S(x)\to\widetilde S_{\rm brems}(x)$ in the numerator,
while in the denominator $\widetilde S(x)\to2\pi\,\delta(x)$ because
the medium-induced change of the cross section is itself of order
$\gamma_\sigma$. With the representation Eq.~(\ref{E10}) we find
\begin{equation}
\langle\Delta \epsilon\rangle_{\rm brems}
=\Gamma_\sigma\int_0^\infty \frac{dx}{2\pi}\,s(x)\,
\frac{F_\beta(x)}{x}\,(e^{-x}-e^{-\beta x}).
\label{E32}
\end{equation}
For the classical limit of hard collisions where $s(x)=1$ this is
\begin{equation}
\langle\Delta \epsilon\rangle_{\rm brems} =\Gamma_\sigma\, \frac{-
7+6\beta+\beta^2+12\ln\beta}{24\pi}.
\label{E34}
\end{equation}

This is to be compared with the average energy transfer by nucleon
recoils, $\langle\Delta \epsilon\rangle_{\rm recoil}=
30\,(\beta-1)\, \beta^{-2}\,T^2/m_N$ \cite{Tubbs}. Therefore, the
ratio between the two is
\begin{eqnarray}
\frac{\langle\Delta\epsilon\rangle_{\rm brems}}
{\langle\Delta \epsilon\rangle_{\rm recoil}}
&=&\frac{\Gamma_\sigma m_N}{T^2}\,\beta^2\, \frac{-
7+6\beta+\beta^2+12\ln\beta}{720\pi\,(\beta-1)}
\nonumber\\ \noalign{\vskip4pt plus1pt}
&=&
\frac{\Gamma_\sigma m_N}{T^2}\left( \frac{1}{36\pi}+ \frac{7(\beta-
1)}{144\pi}+\ldots\,\right).
\label{E35}
\end{eqnarray}
Therefore, the importance of the ``inelastic'' mode of energy transfer
exceeds that of recoils if $\gamma_\sigma>36\pi\,T/m_N$.

We note that the quasi-bremsstrahlung process of Fig.~\ref{Fig1} has a
standard counterpart where neutrino pairs are absorbed or emitted.  We
define a rate of energy transfer in this channel, normalized to the
average neutrino scattering rate in analogy to the above
discussion. By virtue of Ref.~\cite{RSb} the result can be expressed
like Eq.~(\ref{E32}) with $F_\beta(x)= \beta^5 x^5/1440$. The
efficiency of energy transfer relative to recoil effects is
\begin{equation}
\frac{\langle\Delta\epsilon\rangle_{\rm pair}}
{\langle\Delta \epsilon\rangle_{\rm recoil}}
=\frac{\Gamma_\sigma m_N}{T^2}\,\beta^2\, 
\frac{(\beta^2+1)(\beta+1)}{3600\pi}.
\end{equation}
Therefore, the quasi-bremsstrahlung process of Fig.~\ref{Fig1} is 
approximately a factor of 25 more important than pair processes.


\section{Discussion and Summary}

To summarize, we have studied the neutrino-nucleon scattering cross
section taking into account nucleon spin fluctuations. The effect of
random spin fluctuations is to reduce the cross section in a manner
similar to the LPM reduction of photon bremsstrahlung by
multiple-scattering effects. We have derived perturbative results in
terms of a lowest-order spin-density structure function, and also in
terms of the squared matrix element of neutrino-nucleon scattering in
the presence of bystander nucleons or more general external
spin-dependent potentials. In this form, our result agrees with
Sawyer's \cite{Sawyer} related finding.  The low-density limit is
unique unless there are unexpectedly strong spin-spin correlations.

While we have focussed on neutral-current processes, similar
conclusions would obtain for charged-current collisions as stressed in
Refs.~\cite{KJR,Sawyer}.

The explicit low-density results are theoretically interesting, but
their practical significance is limited. It is obvious from
Fig.~\ref{Fig4} that a plausible extrapolation into the high-density
regime vastly differs from the perturbative result for
$\gamma_\sigma=\Gamma_{\sigma}/T\agt 1$ which implies that the
perturbative result cannot be trusted for densities greater than a few
percent nuclear.  While $\Gamma_{\sigma,{\rm OPE}}$ overestimates the
true $\Gamma_\sigma$ at nuclear density, in a SN core one has values
for the true $\gamma_\sigma$ of order a few, perhaps as large as~10
\cite{KJRS,Sigl}. Therefore, the neutrino opacities in the inner SN
core cannot be treated by perturbation theory alone. Near the neutrino
sphere, corresponding to $\gamma_\sigma={\cal O}(1)$, a perturbative
treatment is roughly justified, but the cross-section reduction is
small (a few percent) and thus not overly significant.

Near the neutrino sphere, the most important practical consequence of
nucleon spin fluctuations is likely to be the inelastic or
quasi-bremsstrahlung mode of energy transfer. With Eq.~(\ref{E35}) and
taking $T=5\,\rm MeV$ as a typical neutrino-sphere temperature it is
found to compete with standard recoils for $\gamma_\sigma\agt0.5$. As
this value is representative for conditions around the neutrino
sphere, we confirm that the inelastic mode of energy transfer is about
as efficient as recoils and thus may be important for the formation
of neutrino spectra~\cite{KJRS}.


\section*{Acknowledgments}

We thank Dr.~R.~Sawyer for an illuminating E-mail correspondence
concerning the formalism employed in Ref.~\cite{Sawyer}.  At the
Max-Planck-Institut f\"ur Physik, partial support by the European
Union contract CHRX-CT93-0120 and by the Deutsche
Forschungsgemeinschaft grant SFB 375 is acknowledged. At the
University of Chicago this work was supported by DOE, NSF, NASA, and
the Alexander-von-Humboldt Foundation, at Fermilab by NASA under grant
NAG 5-2788 and by DOE. At Bartol, support by DOE grant
DE-AC02-78ER05007 is acknowledged.



\begin{references}

\bibitem{Sitenko} A.~G.~Sitenko, Electromagnetic Fluctuations in 
     Plasma (Academic Press, New York, 1967). 

\bibitem{SawyerOld} R.~F.~Sawyer, Phys. Rev.~C {\bf 40}, 865 (1989).

\bibitem{pioncondensate} T.~Muto and T.~Tatsumi, Prog. Theor. Phys.
    {\bf 80}, 28 (1988). T.~Muto, T.~Tatsumi, and N.~Iwamoto, Phys.
    Rev. D {\bf 50}, 6089 (1994).

\bibitem{bubblephase} L.~B.~Leinson, Astrophys. J. {\bf 415}, 759
    (1993).

\bibitem{RSb} G.~Raffelt and D.~Seckel, Phys.~Rev.~D  {\bf 52}, 1780
     (1995).

\bibitem{Landau} L.~D.~Landau and I.~Pomeranchuk, Dokl. Akad. Nauk.
     SSSR {\bf 92}, 535 (1953) and ibid.\ {\bf 92}, 735 (1953).
     A. B. Migdal, Phys. Rev. {\bf 103}, 1811 (1956).

\bibitem{Knoll} J.~Knoll and D.~N.~Voskresensky, Phys. Lett. B
     {\bf 351}, 43 (1995).

\bibitem{RSa} G.~Raffelt and D.~Seckel, Phys. Rev. Lett. {\bf 67},
     2605 (1991).

\bibitem{KJR} W.~Keil, H.-T.~Janka, and G.~Raffelt, Phys.~Rev.~D
     {\bf 51}, 6635 (1995).

\bibitem{KJRS} H.-T.~Janka, W.~Keil, G.~Raffelt, and D.~Seckel, Report
     ASTRO-PH/9507023, to be published in Phys. Rev. Lett. (1996).

\bibitem{Sigl} G.~Sigl, Report Fermilab-Pub-95/274-A, to be published
     in Phys. Rev. Lett. (1996).

\bibitem{Sawyer} R.~F.~Sawyer, Phys. Rev. Lett. {\bf 75}, 2260 (1995).

\bibitem{IP} N.~Iwamoto and C.~J.~Pethick, Phys. Rev.~D {\bf 25}, 313
      (1982).

\bibitem{Jackson} J.~D.~Jackson, Classical Electrodynamics, 2nd ed.\
    (John Wiley, New York, 1975).

\bibitem{Raffelt} G.~G.~Raffelt, Stars as Laboratories for Fundamental
    Physics, to be published by the University of Chicago Press
    (1996).

\bibitem{BD} For example J.~D.~Bjorken and S.~D.~Drell,
    Relativistic Quantum Mechanics (McGraw-Hill, New York, 1964).

\bibitem{analyt} R.~J.~Eden, P.~V.~Landshoff, D.~I.~Olive, and 
    J.~C.~Polkinghorne, The Analytic S-Matrix (Cambridge University 
    Press, 1966).  I.~T.~Todorov, Analytic Properties of Feynman 
    Diagrams in Quantum Field Theory (Pergamon Press, Oxford, 1971). 

\bibitem{Tubbs} D.~L.~Tubbs, Astrophys.~J. {\bf 231}, 846 (1979).

\end{references}
\end{document}